\documentclass[aps,prl,floatfix,twocolumn,tightenlines,amsmath,amssymb]{revtex4}

\usepackage{graphicx}
\usepackage{amssymb}
\usepackage{color}
\usepackage{psfrag}
\usepackage{ifsym}
\usepackage{epstopdf}
\usepackage{cancel}

\newcommand{\ket}[1] {| #1 \rangle}

\makeatletter
\renewcommand*\env@matrix[1][c]{\hskip -\arraycolsep
  \let\@ifnextchar\new@ifnextchar
  \array{*\c@MaxMatrixCols #1}}
\makeatother

 {\begin{list}{}%
         {\setlength{\leftmargin}{#1} \setlength{\rightmargin}{#2}}%
         \item[]%
 }
 {\end{list}}

\begin{document}
\title{Photonic Boson Sampling in a Tunable Circuit}

\author
{
Matthew A. Broome$^{1,2,\dagger}$, Alessandro Fedrizzi$^{1,2}$, Saleh Rahimi-Keshari$^{2}$,\\
Justin Dove$^{3}$, Scott Aaronson$^{3}$, Timothy C. Ralph$^{2}$, and Andrew G. White$^{1,2}$\\ 
\normalsize{$^{1}$\emph{Centre for Engineered Quantum Systems},}\\
\normalsize{$^{2}$\emph{Centre for Quantum Computer,and Communication Technology},}\\
\normalsize{School of Mathematics and Physics, University of Queensland, Brisbane 4072, Australia}\\
\normalsize{$^{3}$\emph{Computer Science and Artificial Intelligence Laboratory},}\\
\normalsize{Massachusetts Institute of Technology, Cambridge, MA 02139, USA}\\ 
\normalsize{$^{\dagger}$To whom correspondence should be addressed, m.a.broome@googlemail.com}
}

\begin{abstract}Quantum computers are unnecessary for exponentially-efficient computation or simulation if the Extended Church-Turing thesis---a foundational tenet of computer science---is correct. The thesis would be directly contradicted by a physical device that efficiently performs a task believed to be intractable for classical computers. Such a task is \textsc{BosonSampling}: obtaining a distribution of $n$ bosons scattered by some linear-optical unitary process. Here we test the central premise of \textsc{BosonSampling}, experimentally verifying that the amplitudes of 3-photon scattering processes are given by the permanents of submatrices generated from a unitary describing a 6-mode integrated optical circuit. We find the protocol to be robust, working even with the unavoidable effects of photon loss, non-ideal sources, and imperfect detection. Strong evidence against the Extended-Church-Turing thesis will come from scaling to large numbers of photons, which is a much simpler task than building a universal quantum computer.
\end{abstract}

\maketitle

\noindent Quantum computation has attracted much attention because of the promise of new computational and scientific capabilities. The most famous quantum algorithm is Shor's factoring algorithm \cite{Shors}, which if realised will efficiently factor large composite numbers into their constituent primes, a task whose presumed difficulty is at the basis of the majority of today's public-key encryption schemes. What is not widely appreciated is that the very existence of Shor's algorithm poses a fundamental trilemma: respectively, at least one of the foundational tenets of physics, mathematics, or computer science, is untrue \cite{Trilemma}.

\begin{figure}
  \begin{center}
 \includegraphics[width=1\columnwidth]{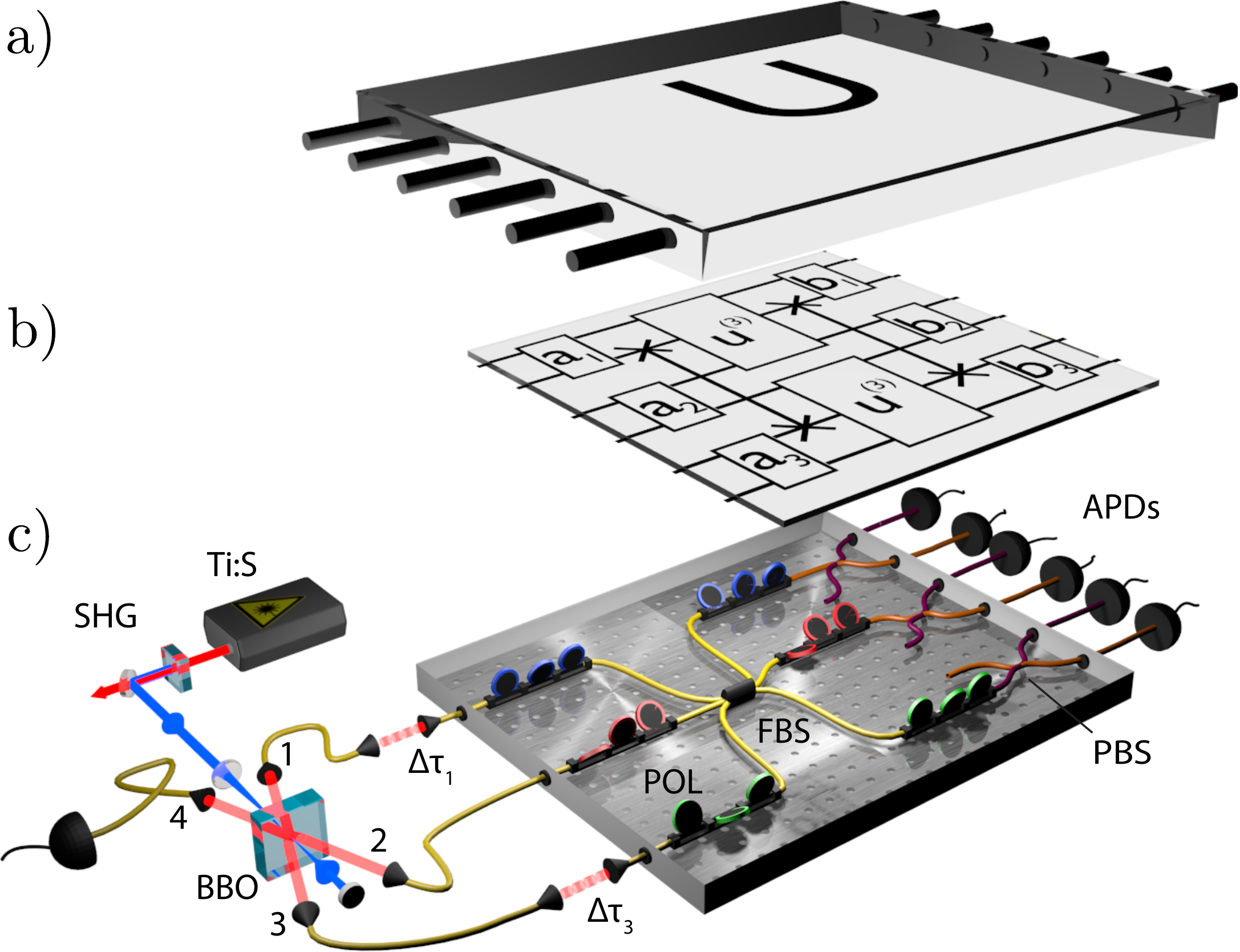}
  \end{center}
\caption{{\bf Experimental scheme for \textsc{BosonSampling}.} (a) In \textsc{BosonSampling} both Alice and Bob must find the output distribution from some unitary, $U$, for multi-boson inputs; Alice \& Bob respectively possess classical and quantum resources. (b) The equivalent circuit diagram of our unitary. Three spatial mode inputs each encode two orthogonal polarisation modes which can be arbitrarily combined by the unitaries $a_1$, $a_2$ and $a_3$. A $3{\times}3$ unitary evolution, $u^{(3)}$, interferes all three modes of the same polarisation, these are recombined at the output by $b_1$, $b_2$ and $b_3$. Note the swap gates between modes 2 and 5, which reflect that only modes of the same polarisation interact in $u^{(3)}$. (c) Experimental schematic. Two pairs of single photons are produced via spontaneous parametric downconversion in a nonlinear crystal (BBO), driven by a frequency doubled (SHG) femtosecond laser (Ti:S) see section I of Supplementary Material. Up to three of the four photons are injected into the \textsc{BosonSampling} circuit; photons 1 and 3 can be delayed or advanced with respect to photon 2 by $\Delta \tau_{1}$, $\Delta \tau_{3}$ respectively; the fourth photon serves as a trigger. The \textsc{BosonSampling} circuit is constructed from a combination of $3{\times}3$ and $2{\times}2$ integrated optical fibre beam-splitters. The local unitaries, $a_1... b_3$ are implemented with polarisation controllers (POL); $u^{(3)}$ is implemented by a biased $3{\times}3$ non-polarising tritter (FBS), its outputs are mapped to 6 spatial modes by three polarising fibre beam-splitters (PBS). The outputs of the final $6{\times}6$ circuit are coupled to single photon avalanche diodes (APDs) whose signals are processed by a counting logic based on a field-programmable gate-array circuit.}
  \label{fig:setup}
\end{figure}

Shor's algorithm states that efficient factoring can be done on a quantum computer, which is thought to be a realistic physical device. It may be that a scalable quantum computer is not realistic, if for example quantum mechanics breaks down for large numbers of qubits \cite{HoloDavies}. If, however a quantum computer is a realistic physical device at all scales, then the Extended Church-Turing thesis---that any computational function on a realistic physical device can be \emph{efficiently} computed on a probabilistic Turing Machine---means that a classical, efficient, factoring algorithm exists. Such an algorithm, long sought-after, would enable us to break public-key cryptosystems like RSA. A third possibility is that the Extended Church-Turing thesis itself is wrong.

How do we answer this trilemma? As yet there is no evidence that quantum mechanics doesn't apply for large-scale quantum computers---that will need to be tested directly via experiment---and there is no efficient classical factoring algorithm or mathematical proof of its impossibility. This leaves examining the validity of the Extended Church-Turing thesis. One approach would be to take a task that was believed, on strong evidence, to be intractable for classical computers, and to build a physical device that performs the task efficiently. This would directly contradict the Extended Church-Turing thesis. 

One such task is \textsc{BosonSampling}: obtaining a representative sample-distribution for $n$ bosons scattered by some linear-optical unitary process, given by a $m{\times}m$ matrix $U$ \cite{AA1,AA2}. This task likely becomes intractable for large $n$, for reasons related to the fact that the amplitudes of $n$-boson processes are given by the permanents of $n{\times}n$ sub-matrices of $U$, and calculating the permanent is a so-called `\#P-complete' problem~\cite{Valiant1979}---a complexity class above even `NP-complete'. Note that \textsc{BosonSampling} \emph{itself} is not thought to be \#P-complete: the ability to solve it lets us sample random matrices most of which have large permanents, but probably does not let us estimate the permanent of a particular, given matrix. However, by using the fact that the permanent is \#P-complete, Ref.~\cite{AA1} showed that even for the `easier' task of \textsc{BosonSampling}, any fast classical algorithm would lead to drastic consequences in classical computational complexity theory, notably collapse of the `polynomial hierarchy'. (A similar task, instantaneous quantum polytime, IQP, was recently proposed in Ref.\cite{Bremner08022011}: experimentally it appears far more difficult to achieve).

Here we test the central premise of \textsc{BosonSampling} in practice, experimentally verifying that the amplitudes of $n{=}3$ photon scattering events are given by the permanents of $n{\times}n$ sub-matrices of the unitary operator $U$ describing the photonic quantum computer. We find the protocol to be robust, working even with the unavoidable real-world effects of photon loss, non-ideal photon sources, and imperfect photon detection. We make use of a novel method for experimental characterisation of near-unitary evolutions which is both fast and accurate \cite{Saleh}.

Imagine a race between two participants, Alice, who only possesses classical resources, and Bob, who possesses quantum resources. They are given some unitary, $U$, and agree on a specific $n$-boson input configuration. Alice calculates an output sample-distribution with a classical computer; Bob either builds---or programs an existing---linear-photonic network,  sending $n$ single-photons through it and obtaining his sample by measuring the output distribution. The race ends when both return samples from the distribution. In principle the validity of Alice and Bob's samples can only be established by individually comparing them against knowledge of the full distribution; in practice they can check for consistency by seeing if their distributions compare to within error. The winner is whoever returns a sample fastest. 

\begin{figure*}
  \begin{center}
 \includegraphics[width=\textwidth]{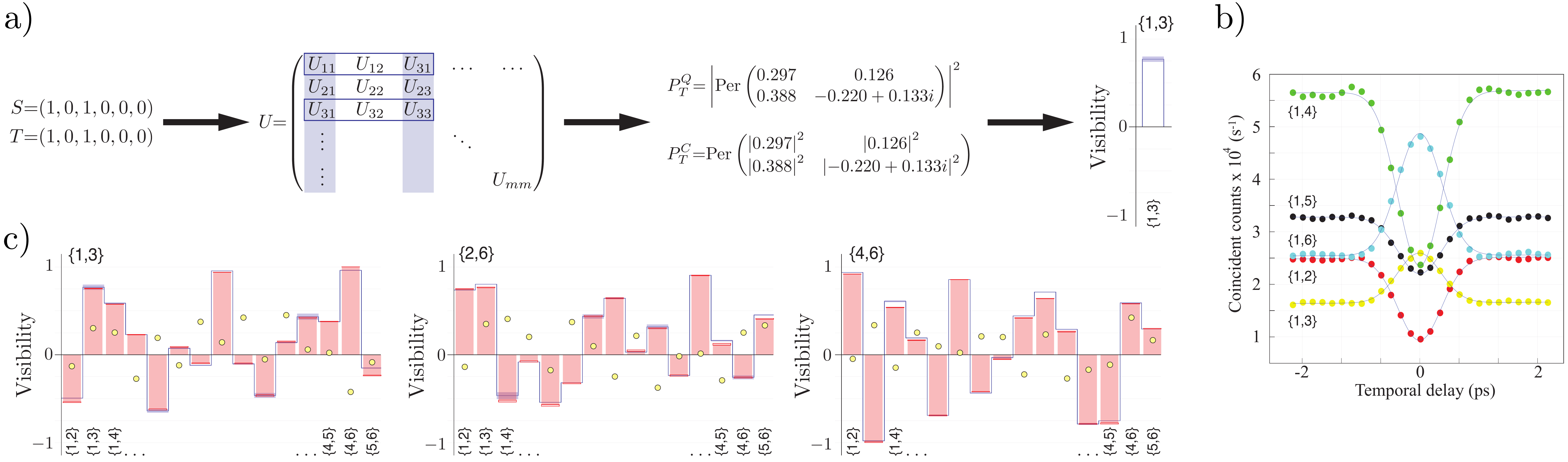}
  \end{center}
\caption{{\bf Two-photon \textsc{BosonSampling}.} (a) Outline of the technique Alice uses to predict the visibility from the unitary evolution $U$. (We present the fully characterised unitary $U$ in the section II of the Supplementary Material as well as an example permanent calculation). Alice's prediction is shown by the blue-line envelope at far right; the light-blue box represents the uncertainty, obtained by 10 separate characterisations of the unitary. (b) Sample two-photon quantum interferences: the five output combinations $\{1,m\}$ for the input configuration of $\{1,5\}$. Errors are smaller than marker size and the solid blue lines are Gaussian fits used to calculate the visibility from Eq.~\ref{eq:vis}. (c) Experimental data showing Alice's (predicted) and Bob's (measured) visibilities for two-photon quantum interference. The input configurations are shown in the top left of each panel; the output modes are labelled at the bottom of the plot. The solid blue-line envelopes are Alice's predictions for visibility based on her measurement of U; the orange bars are Bob's measured visibilities; the yellow circles are the visibility predictions if coherent input-states were used instead of two-photon inputs. Errors are given by the light-blue and dark-red boxes at the extrema of each data set. \vspace{-5mm} } 
  \label{fig:results1}
\end{figure*}

As $n$ becomes large, it is conjectured that Bob will always win, since Alice's computation times increases exponentially, whereas Bob's experimental time does not. It becomes intractable to verify Bob's output against Alice's, and---unlike the case of Shor's algorithm---there is no known efficient algorithm to verify the result~\cite{AA1}. Importantly, however, one can take a large instance---large enough for verification via a classical computer---and show that Bob's quantum computer solves the problem much faster, thereby strongly suggesting that the same behaviour will continue for larger systems, casting serious doubt on the Extended Church-Turing Thesis.

In a fair race, Bob will verify that his device actually implements the target unitary. An alternative fair version of the race is to give both Alice and Bob the same physical device---instead of a mathematical description---and have Alice characterise it before she predicts output samples via classical computation. Alice can use a characterisation method that neither requires nonclassical resources nor adds to the complexity of the task \cite{Saleh}.

We performed our \textsc{BosonSampling} demonstration in an optical network with $m{=}6$ input and output modes, and $n{=}2$ and $n{=}3$ photon inputs. We implement a randomly-chosen unitary which is fully-connected, i.e. every input is distributed to every output. The experimental setup is shown in Fig.~\ref{fig:setup}. The 6-input${\times}$6-output modes of the unitary $U$ are represented by two orthogonal polarisations in $3{\times}3$ spatial modes of a fused-fibre-beamsplitter, an intrinsically stable and low-loss device. The mode mapping is  $\{1,...,6\}{=}\{\ket{H}_1,\ket{V}_1,\ket{H}_2,\ket{V}_2,\ket{H}_3,\ket{V}_3\}$, where $\ket{H}_1$ is the horizontally polarised mode for spatial mode $1$. We use polarisation controllers at the inputs and outputs to modify the unitary, see the equivalent circuit diagram in Fig~\ref{fig:setup}B). 

Previously, quantum photonic circuits have been characterised via quantum process tomography \cite{Obrien:protomo}, requiring an exponentially-increasing number of measurements. A recent theoretical proposal---yet to be experimentally realised---suggests using a linear number of non-classical interferences \cite{laing2012sst}, however the output coincidence signal falls off quadratically with circuit size. Here we use an efficient, method for this challenging task which requires just $(2m{-}1)$ measurement combinations of single- and dual-mode coherent-state inputs to the $m{\times}m$ network. The photonic network is described by a mapping between the input, $a^{\dagger}_i$, and output, $a^{\dagger}_j$,  creation operators,
\begin{equation}
a^{\dagger}_j=\sum^m_i U_{i,j}a^{\dagger}_i \vspace{-1mm} 
\label{eq:unitary}
\end{equation}
where $U_{i,j}{=}r_{ij}e^{i\theta_{ij}}$. In brief, Alice can determine the moduli $r_{ij}$, by preparing single mode coherent states into modes $1...m$ and measuring the output intensities, which are proportional to the occupation numbers $n_{i1}...n_{ij}...n_{im}$, \vspace{-1mm}
\begin{equation}
r_{ij}=\sqrt{n_{ij}}. \vspace{-1mm}
\end{equation}
Next she determines the phases $\theta_{ij}$ by preparing pairwise interferometrically stable dual-mode coherent states between outputs $1$ and all other outputs $2...m$. Further details, including a full theoretical derivation, can be found in ref.~\cite{Saleh}.

Having obtained $U$, Alice calculates the probability of bosonic scattering events in the following way~\cite{Scheel2004,AA1}. Given the input and output configurations $S{=}(s_1,...,s_m)$ and $T{=}(t_1,...,t_m)$ with boson occupation numbers $s_i$ and $t_j$ respectively, she produces an $n{\times}m$ matrix $U_{T}$ by taking $t_j$ copies of the $j^{\textrm{th}}$ column of $U$. Then, she forms the $n{\times}n$ submatrix $U_{ST}$ by taking $s_i$ copies of the $i^{\textrm{th}}$ row of $U_{T}$. The probability for the scattering event $T$, for indistinguishable input photons $S$, is given by $P^{Q}_{T}{=}\left|\textrm{Per}(U_{ST})\right|^2$. Conversely, the classical scattering probabilities---when the input photons are distinguishable---is given by $P^{C}_{T}{=}\textrm{Per}(\tilde{U}_{ST})$, where $\tilde{U}_{ST_{ij}}{=}\left|{U}_{ST_{ij}}\right|^2$.

\begin{figure*}
  \begin{center}
 \includegraphics[width=1\textwidth]{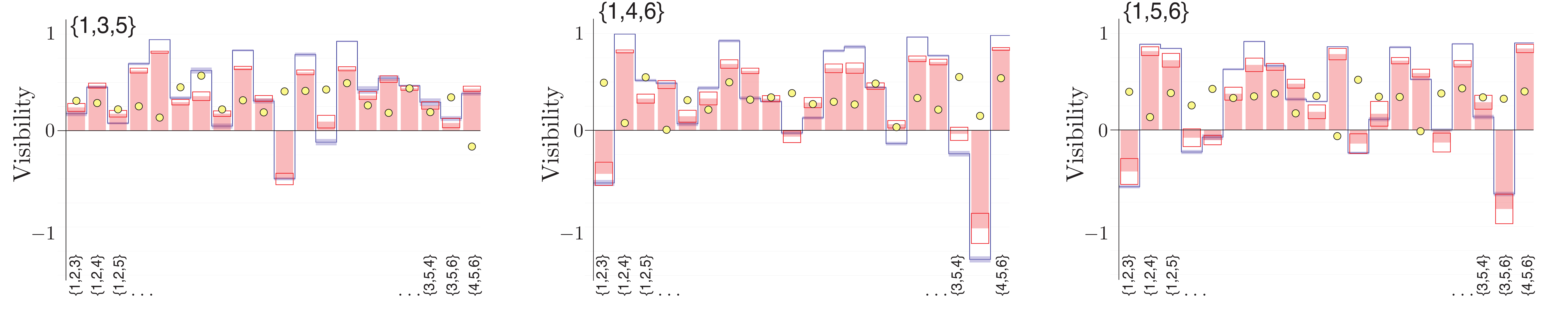}
  \end{center}
\caption{{\bf Three-photon \textsc{BosonSampling}.} Experimental data showing Alice's (predicted) and Bob's (measured) visibilities for three-photon quantum interference. The input configurations are shown in the top left of each panel; the output modes are labelled at the bottom of the plot. The solid blue-line envelopes are Alice's predictions for visibility based on her measurement of U; the orange bars are Bob's measured visibilities; the yellow circles are the visibility predictions if coherent input-states were used instead of three-photon inputs. Errors are given by the light-blue and dark-red boxes at the extrema of each data set. }
  \label{fig:results2}
\end{figure*}

\vspace{1mm} Bob on the other hand experimentally prepares the $n$-photon Fock state $\left|t_1,...,t_m\right>$. After injecting the desired input to the circuit, he determines the probability of the scattering event $T$ by projecting onto its corresponding state using single-photon detectors connected to a coincidence counting logic, see Fig.~\ref{fig:setup}C). 

We prepare near-single-photon Fock states via spontaneous parametric downconversion in a nonlinear uni-axial crystal, Fig.~\ref{fig:setup}C), see section I in Supplementary Material for detail. Once the photons pass through the network, they are detected by single-photon avalanche diodes. The \textsc{BosonSampling} protocol measures the frequency of output events, i.e. raw coincident photon counts. 

These however are strongly affected by differences in efficiency between photon counters. To remove this effect we measure the non-classical interference visibility,
\begin{equation}
V_T=\frac{P^{C}_{T}-P^{Q}_{T}}{P^{C}_{T}},
\label{eq:vis}
\end{equation}
where $P^{Q}_{T}$ and $P^{C}_{T}$ are the quantum and classical probabilities for the output configuration $T$ measured for completely indistinguishable and distinguishable photons respectively. Distinguishable statistics are obtained by introducing a temporal delay, $\Delta \tau$, between the input photons. When all photons are delayed by significantly more than their respective coherence lengths, $L$, true two-photon quantum interference cannot occur. Figure~\ref{fig:results1}A) outlines the technique Alice uses to predict the visibility from the unitary evolution $U$.

For $n{=}2$, the output $T$ is monitored continuously as a function of the temporal delay between the two input photons; as is typically done with the well-known Hong-Ou-Mandel effect \cite{PhysRevLett.59.2044}. For $n{=}3$, however, the low four-photon count rates mean that such a measurement will be degraded due to inevitable optical misalignment and drift encountered over necessarily long experimental runtimes. Therefore, for $n{=}{3}$ the probabilities $P^{C}_T$ are obtained from just two measurement settings, $P^{C}_{T}(1){=}\{-\Delta\tau_\infty,0,\Delta\tau_\infty\}$ and  $P^{C}_{T}(2){=}\{\Delta\tau_\infty,0,-\Delta\tau_\infty\}$, where $\{\tau_1,\tau_2,\tau_3\}$ are the temporal delays of photons 1, 2 and 3 with respect to photon 2, and $\Delta\tau_\infty{\gg}L/c$. $P^{C}_{T}$ is calculated as the average of these two probabilities to account for optical misalignment. Accordingly, $P^{Q}_{T}$ are obtained with a single measurement of the output frequencies for completely indistinguishable photons, given by the delays $\{0,0,0\}$.

As a first test of our system we compare Alice's and Bob's methods by injecting $n{=}2$ photons into the \textsc{BosonSampling} circuit. Figure~\ref{fig:results1}B) shows a representative set of non-classical two-photon interference patterns as a function of the temporal delay, $\Delta\tau$. A more complete picture is given in Fig.~\ref{fig:results1}C), here we obtain sample distributions for three combinations of input configurations and measure at all $C(6,2)$ (6 choose 2) possible output configurations. We compare Alice's and Bob's measurements using the average $L_{1}$-norm distance per output configuration, $\mathcal{L}_1{=}\frac{1}{C(m,n)} \sum_T\left|V^A_T{-}V^B_T\right|$; here it is $\mathcal{L}_1{=}0.027$, showing excellent agreement between them.

Next we show that if Alice uses her available resources---notably, coherent-states---to perform an analogous experiment to Bob's she will not obtain the same results (see section III in Supplementary Material). Her coherent-state predictions---given by the yellow circles in Fig.~\ref{fig:results1}C)---are clearly different to Bob's quantum measurements, with $\mathcal{L}_1{=}0.548$. This large disagreement indicates that the quantum distribution is sufficiently different from the classical distribution---so that this unitary isn't some special choice that is easy to sample classically---and highlights that Bob is accurately sampling from a highly nonclassical distribution.

 \begin{figure}
\begin{center}
 \includegraphics[width=1\columnwidth]{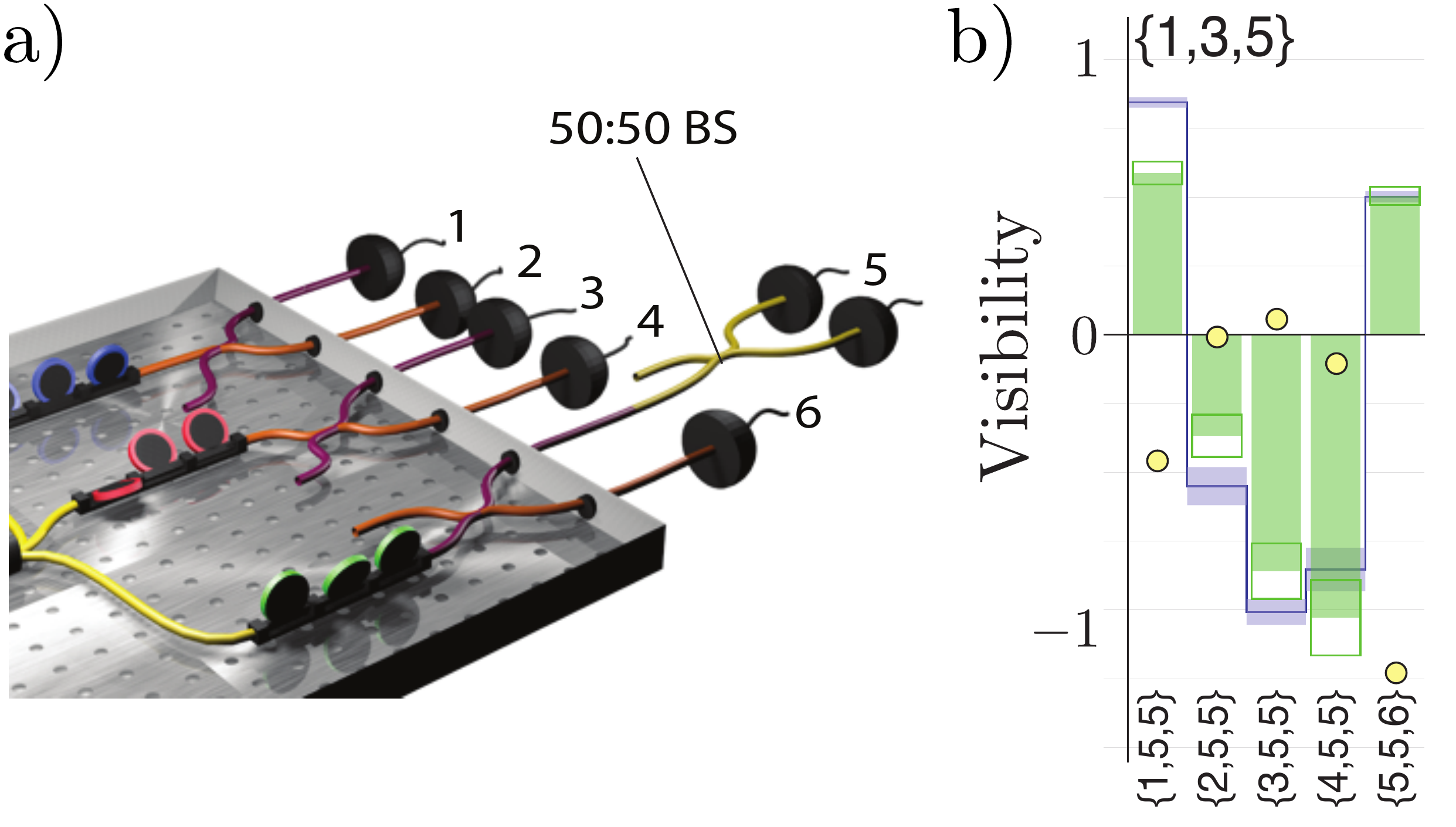}
  \end{center}
\caption{{\bf Three-photon \textsc{BosonSampling} with colliding outputs.} (a) Number resolution was achieved with a $50{:}50$ fibre beam-splitter at the output of mode $5$ and an additional single-photon detector. Note that an imperfect splitting ratio for this fiber beam-splitter impedes only the effective efficiency of our number resolving scheme~\cite{PhysRevLett.76.2464,PhysRevA.63.033812}. (b) For an input configuration of $\{1,3,5\}$, and measuring two-photons in output mode $5$, the solid blue-line envelope shows Alice's predictions; the green bars are Bob's measured visibilities; and the yellow circles are the visibility predictions if coherent input-states were used instead of three-photon inputs. Errors are given by the light-blue and green boxes at the extrema of each data set.}
  \label{fig:results4}
\end{figure}

Figure \ref{fig:results2} shows the results for $n{=}3$ \textsc{BosonSampling}. Here we find $\mathcal{L}_1{=}0.122$: we attribute the larger average distance chiefly to the increased ratio of higher-order photon emissions in the three-photon experiment compared with the two-photon case (see section IV in Supplementary Material). Again, Alice's efficiently-computed coherent-state predictions are clearly different to Bob's measurements, with $\mathcal{L}_1{=}0.358$. 

Having tested all possible `non-colliding'  output configurations---that is, 1-photon per output-mode---we also tested `colliding' configurations with up to 2-photons per output-mode. This requires photon-number resolution~\cite{PhysRevLett.76.2464,PhysRevA.63.033812}, using the method shown in Fig.~\ref{fig:results4}A). Figure~\ref{fig:results4}B) shows once again good agreement between Alice's predicted and Bob's measured sample distributions, $\mathcal{L}_1{=}0.153$, and a much larger distance between Bob's  measurements and Alice's coherent-state predictions, $\mathcal{L}_1{=}0.995$.  

\begin{figure}
  \begin{center}
 \includegraphics[width=0.8\columnwidth]{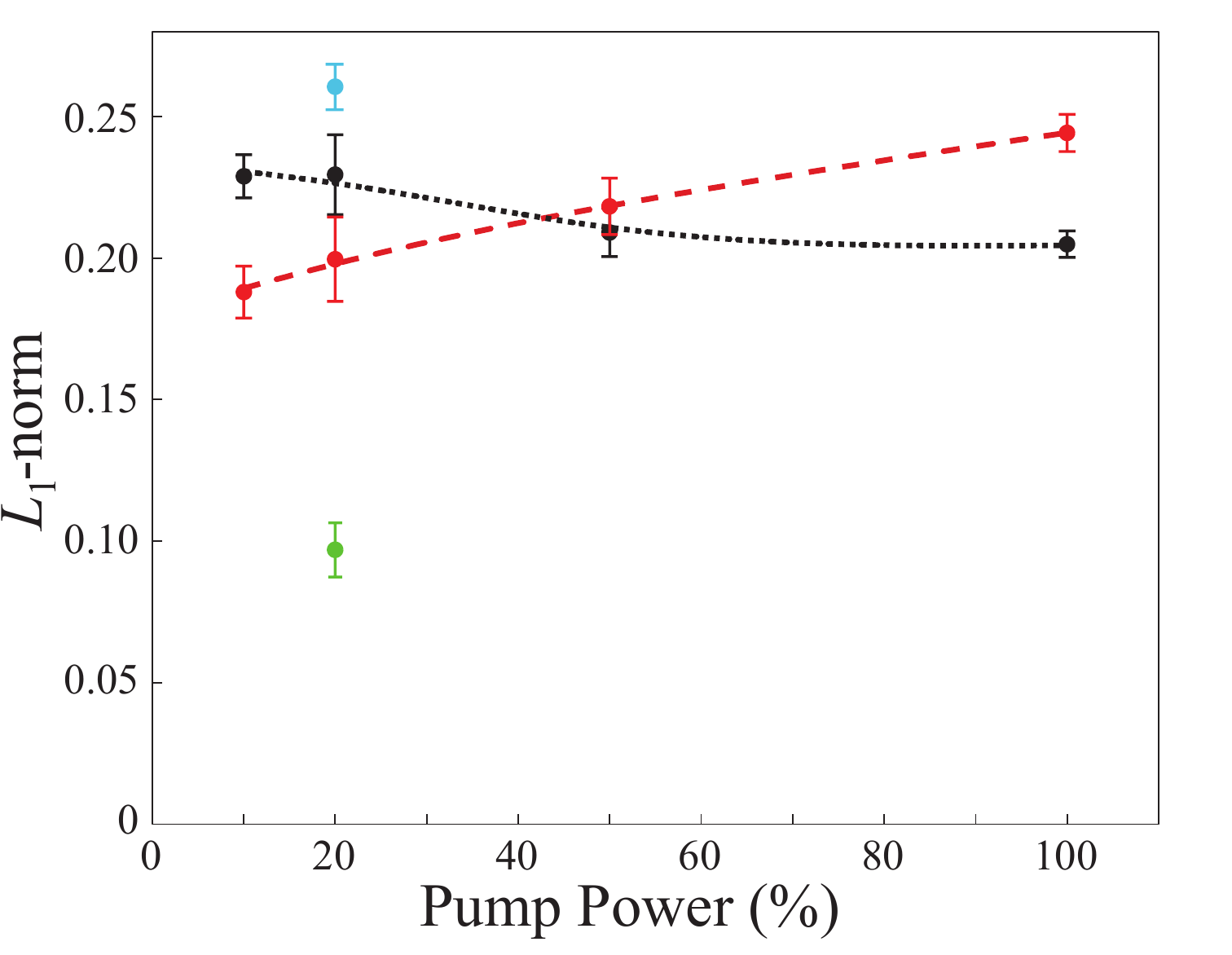}
  \end{center}
\caption{{\bf Imperfect Fock states in \textsc{BosonSampling}.} For an input configuration of $\{1,3,5\}$, increasing the downconversion pump power introduces larger proportions of higher-order photon numbers. \emph{red dashed-line and data}: The $L_{1}$-norm distance between Alice's predictions for three-photon inputs and Bob's measured visibilities.  \emph{black dotted-line and data}: The $L_{1}$-norm distance between Alice's predictions for coherent-state inputs and Bob's measured visibilities. The lines are best fits to guide the eye. Data shown in black and red were taken using a spectral filter on the downconverted modes with FWHM of $5$~nm. Decreasing the spectral filter to $2$~nm significantly reduces the $L_{1}$-norm distance between Alice's predictions for Fock-state inputs and Bob's measured visibilities (green), while increasing the distance between Bob's measurements and those predicted using coherent state inputs (blue).}
  \label{fig:5}
\end{figure}

Alice's Fock-state predictions and Bob's measurement results do not quite overlap to within error for $n{=}2$, and more notably disagree for $n{=}3$. This indicates the presence of respectively small and moderate systematic differences between Alice's and Bob's methods for obtaining non-classical visibilities. This is as expected, since Alice's calculations are for indistinguishable Fock-state inputs, and Bob does not actually have these. It is well-known that the conditioned outputs from downconversion have higher-order terms, i.e. a small probability of producing more than one photon per mode (see section IV of the Supplementary Material), and are also spectrally-entangled, leading to a small degree of distinguishability. The higher-order terms increase with source brightness, as Fig.~\ref{fig:5} shows, this increases the distance between Alice's Fock-state predictions and Bob's measurements, pushing the latter into a more classical regime. Thus when using downconversion, source brightness must be kept low, but since downconversion is probabilistic its brightness decreases exponentially with $n$: the best demonstrations to date have been $n{=}8$ at a rate of ${\sim}10^{-3}$ Hz \cite{yao2012observation}.  Some gains can be made by increased spectral filtering, which decreases photon distinguishability. As the top and bottom data points in Fig.~\ref{fig:5} show, this significantly moves Bob's measurements towards Alice's Fock-state predictions and away from the classical case. 

Strong evidence against the Church-Turing thesis will come from demonstrating \textsc{BosonSampling} with a larger-sized system where Bob's experimental sampling is far faster than Alice's calculation and where classical verification is still barely possible---according to \cite{AA1}, this regime is on the order of $n{=}$20 to $n{=}$30 photons in a network with with $m{\gg}n$ modes. 

This is beyond current technologies, but rapid improvements in efficient detection, low-loss circuitry and improved photon sources are highly promising. Highly-efficient, photon-number-resolving, detectors are now well-developed: e.g. transition edge sensors \cite{lita2008counting} have high intrinsic efficiencies, ${>}95\%$, recently enabling whole-system detection efficiencies with entangled-photon pairs of ${>}60\%$ \cite{smith2012cqs}. Integrated circuitry is also being rapidly developed. Nonclassical interference of photon pairs has been observed in periodically-bounded \cite{owens2011two} and linear \cite{peruzzo2010qwc} waveguide arrays, up to $m{=}21$. Whilst not strictly necessary, reconfigurable integrated circuits \cite{shadbolt2012generating,metcalf2012multi} could be used to sample from an ensemble of different unitaries. However, if they are to be used for high-$n$ demonstrations, these circuits will have to be significantly improved to reduce their very high losses, e.g. incurred by coupling photon sources to waveguide arrays. Finally, triggered or heralded photon-sources will be required: Fig.~\ref{fig:5} makes clear that downconversion alone is not a suitable photon source. Developing better photon sources is a focus of research worldwide, again with promising recent developments, e.g. quantum-dot photon sources with production efficiencies of ${>}40\%$ \cite{dousse2010ultrabright}, orders-of-magnitude higher than downconversion. \textsc{BosonSampling} could also be demonstrated in other engineered quantum technologies, such as  superconductors, neutral atoms, or, in an attractive recent proposal, trapped ions: the phononic modes of the ions are the bosons; the ions act as the programmable beamsplitters of the phononic network \cite{PhysRevA.85.062329}. 

An important open question remains as to the practical robustness of large implementations. Unlike the case of universal quantum computation, there are no known error correction protocols for  \textsc{BosonSampling}, or indeed any of the models of intermediate quantum computation, such as deterministic quantum computing with one qubit (DQC1) \cite{PhysRevLett.100.050502, PhysRevLett.101.200501}, temporally unstructured quantum computation (IQP) \cite{Bremner08022011}, or permutational quantum computing (PQC) \cite{jordan2010permutational}. These intermediate models have garnered much attention in recent years due both to the inherent questions they raise about quantum advantage in computing, and because some of them can efficiently solve problems believed to be classically intractable, e.g. DQC1 has been applied in fields that range from knot theory \cite{shor2008estimating} to quantum metrology \cite{PhysRevA.77.052320}. Our experimental results are highly promising with regard to the robustness of \textsc{BosonSampling}, finding good agreement even with clearly imperfect experimental resources; equally heartening, a recent theoretical study provides evidence that photonic \textsc{BosonSampling} retains its computational advantage even in the presence of loss~\cite{PhysRevA.85.022332}.

\subsection*{Acknowledgments}
We thank Robert Fickler for help with characterisation, Marcelo de Almeida, Devon Biggerstaff, and Geoffrey Gillett for experimental assistance, and Alex Arkhipov, Michael Bremner, and Terry Rudolph for discussions. This work was supported in part by: the Australian Research Council's Federation Fellow program (FF0668810), Centre for Engineered Quantum Systems (CE110001013), and Centre for Quantum Computation and Communication Technology (CE110001027); the University of Queensland Vice-Chancellor's Senior Research Fellowship program; the National Science Foundation's Grant No. 0844626, and a Science and Technology Centre grant; a DARPA Young Faculty Award grant; a TIBCO Chair; and a Sloan Fellowship.

\newpage

\begin{center}
{\LARGE Experimental Boson Sampling: Supplementary Material
}
\end{center}
\vspace{10mm}

\noindent{\bf I. Four-photon source and \textsc{BosonSampling} circuit}\vspace{5mm}

\noindent A mode-locked Ti:Sapphire (\textit{Coherent 900 HP}) laser with a repetition rate of $76$~MHz, $100$~fs pulses and an average output power of ${\sim}3.8$~W at $820$~nm is frequency doubled in a $2$~mm long bismuth borate (BiBO) nonlinear crystal to give ${\sim}1.5$~W centred at $410$~nm. Photon-pairs are produced via spontaneous parametric downconversion: the two pairs are produced by forward and backward passes of a $2$~mm long beta-barium-borate (BBO) nonlinear crystal cut for type-I phase matching. Single photons pass through spectral filterers (FWHM $5$~nm) before being coupled into single mode optical fibres. At $100\%$ pump power the forward and backward passes of the source produce approximately $290$~kHz and $180$~kHz two-fold coincidences, with a maximum four-fold rate of $1.20$~kHz. The discrepancy in two-fold coincidences between forward and backward passes is due to differing focussing conditions in each case.

Single photons are directed into the \textsc{BosonSampling} circuit using a combination of calcite beam-displacers and waveplates  The forward pass of the downconversion source is used for the two-photon measurements and the source is run at $20\%$ of the maximum pump power to reduce the ratio of higher-order photon emission. With an average input coupling efficiency of $64\%$ into all modes of the circuit, we obtain an average two-photon coincidence rate of $260$~Hz across all modes. When injecting three photons into the circuit, the remaining fourth photon is sent directly to a single photon detector to serve as a trigger for its twin. Again, running at $20\%$ maximum pump power we obtain an average four-fold coincidence rate of $185$~mHz. For all measurements in the main text we use an additional $2$~nm filter at the input of the \textsc{BosonSampling} circuit, this reduces the effect of group-velocity mismatch between interfering photons from independent downconversion events.

\vspace{5mm}
\noindent{\bf II. Measured unitary matrix and calculation of permanents \vspace{5mm}}

In this section we present a sample measured unitary evolution and demonstrate how we generate sub-matrices whose permanents are used to calculate the probability amplitudes of bosonic scattering events. 

Changes in laboratory conditions cause slight changes to the in the optical fibres network we use as our \textsc{BosonSampling} circuit. For this reason we measure the evolution $U_{\textrm{exp}}$, of the linear optical network after each of Bob's experimental runs, ensuring that Alice generates the most representative scattering probabilities. Below we give the unitaries that were measured after we completing the two- and three-photon non-classical interference measurements respectively,

\begin{footnotesize}
\begin{widetext}
\begin{equation}
U^{2\textrm{photon}}_{\textrm{exp}}=
\begin{pmatrix}[r]
0.297 	 &0.325+0.000i 	 &0.126+0.000i 	 &0.500+0.000i 	 &0.430+0.000i 	 &0.253+0.000i 	\\ 
0.330 	 &-0.302-0.011i 	 &0.001+0.503i 	 &0.028-0.390i 	 	&0.221+0.118i 	 	&-0.385-0.213i 	 \\ 
0.388 	 &0.182+0.248i 	 &-0.220+0.133i 	 &-0.212+0.204i 	 &-0.127-0.386i 	 &0.108-0.081i 	 \\ 
0.311 	 &-0.220-0.315i 	 &-0.169-0.246i 	 &0.190+0.157i 	 &-0.073-0.089i 	 &-0.227+0.355i 	 \\ 
0.396 	 &-0.222-0.169i 	 &0.387-0.130i 	 	&-0.265+0.004i 	 &-0.103+0.202i 	 &0.353-0.112i 	 \\ 
0.279 	 &0.322+0.244i 	 &-0.101-0.239i 	 &-0.051-0.400i 	 &-0.184+0.320i 	 &-0.217+0.074i 	 \nonumber
\end{pmatrix}
\end{equation}
\vspace{5mm}
\begin{equation}
U^{3\textrm{photon}}_{\textrm{exp}}=
\begin{pmatrix}[r]
 0.334 	 & 0.277+ 0.000i 	 & 0.125+ 0.000i 	 & 0.479+ 0.000i 	 & 0.415+ 0.000i 	 & 0.237+ 0.000i 	\\ 
 0.273 	 &-0.329-0.051i 	 & 0.055+ 0.478i 	 & 0.021-0.121i 	 & 0.197+ 0.128i 	 &-0.345-0.253i 	\\ 
 0.420 	 & 0.140+ 0.242i 	 &-0.191+ 0.198i 	 &-0.195+ 0.204i 	 &-0.139-0.393i 	 & 0.113-0.085i 	\\ 
 0.284 	 &-0.197-0.367i 	 &-0.194-0.224i 	 & 0.189+ 0.190i 	 &-0.072-0.106i 	 &-0.278+ 0.333i 	\\ 
 0.340 	 &-0.329-0.049i 	 & 0.328-0.312i 	 &-0.144+ 0.042i 	 &-0.131+ 0.187i 	 & 0.283-0.216i 	\\ 
 0.324 	 & 0.344+ 0.036i 	 &-0.114-0.101i 	 &-0.206-0.398i 	 &-0.111+ 0.351i 	 &-0.098+ 0.208i 	\nonumber
\end{pmatrix}
\end{equation}
\vspace{5mm}
\end{widetext}
\end{footnotesize}

\begin{normalsize}
For our example calculation we will input photons with the configuration $S{=}(1,0,1,0,0,0)$ and look at the output modes given by $T{=}(0,1,0,0,1,0)$. First we generate the sub-matrix $U_{T}$ by selecting $t_j$ copies of the $j^{\textrm{th}}$ column of $U^{2\textrm{photon}}_{\textrm{exp}}$~\cite{Scheel2004,AA1} to give,
\end{normalsize}

\vspace{5mm}
\begin{small}
\begin{equation}
U_{T}=
\begin{pmatrix}[r]
0.325+0.000i 	 &0.430+0.000i \\
-0.302-0.011i  	 &0.221+0.118i \\ 
0.182+0.248i	 &-0.127-0.386i \\ 
-0.220-0.315i	 &-0.073-0.089i \\ 
-0.222-0.169i  	 &-0.103+0.202i\\ 
0.322+0.244i 	 &-0.184+0.320i 
\end{pmatrix}.
\end{equation}
\vspace{5mm}
\end{small}

\hspace{-2mm}Then we generate $U_{ST}$ by selecting $s_i$ copies of the $i^{\textrm{th}}$ row of $U_T$ to give

\vspace{5mm}
\begin{small}
\begin{equation}
U_{ST}=
\begin{pmatrix}[r]
0.325+0.000i 	 &0.430+0.000i \\
0.182+0.248i	 &-0.127-0.386i 
\end{pmatrix}.
\end{equation}
\vspace{5mm}
\end{small}

\hspace{-2mm}The probabilities of observing the even $T$ for completely indistinguishable and distinguishable photons are given by $P^Q_T$ and $P^C_T$ respectively,
\begin{eqnarray}
P^Q_T &=& \left|\textrm{Per}\left(U_{ST}\right)\right|^2  \\
&=& 0.0017\nonumber \\
P^C_T &=& \textrm{Per}\left(\tilde{U}_{ST}\right) \\
&=& 0.0349 \nonumber
\end{eqnarray}
where $\tilde{U}_{ST_{ij}}{=}\left|U_{ST_{ij}}\right|^2$. The above probabilities give a predicted non-classical interference visibility of $V{=}(P^C_T-P^Q_T)/P^C_T{=}0.951$, which compares well to Bob's observed value of 0.939 for this input/output configuration.

The estimated errors for predicted visibilities given in the main text were calculated as the standard deviation of the predicted visibilities from 10 separate unitary characterisations.

\vspace{5mm}
\noindent{\bf III. Calculation of visibility using coherent-state inputs \vspace{5mm}}

\noindent We calculate interference visibilities by injecting $n$ equal-amplitude coherent states with normalised electric field amplitudes $E_i{=}e^{i\theta_i}$, into the input modes $\{i,...,m\}$ of the linear optical network. The input vector ${\bf E}{=}(E_1,...,E_i...,E_n)$ is transformed under $U$ such that the electric field at output mode $j$ is given by
\begin{equation}
E_j=\sum_i U_{ij}E_i.
\end{equation}
When the input coherent states overlap with a zero time delay the $2n$-order correlation in electric field between detectors at output modes $\{j,...,m\}$ is given by the phase averaged cross-correlation function~\cite{metcalf2012multi,PhysRevA.28.929},
\begin{equation}
P^{(0)}=\frac{1}{(2\pi)^n} \int^{2\pi}_0 \hdots  \int^{2\pi}_0 \prod_{j=1}^m\left|E_j\right|^2 d\theta_i \hdots d \theta_n.
\end{equation}
Conversely, if input coherent states are delayed by significantly more than their coherence lengths no interference can occur between them, leaving the cross-correlation function as an incoherent sum of input fields at the output modes $j$,
\begin{equation}
P^{(\infty)}=\frac{1}{(2\pi)^n} \int^{2\pi}_0 \hdots  \int^{2\pi}_0 \prod_{j=1}^m\left(\sum_i \left|E^{(i)}_j\right|^2\right) d\theta_i \hdots d \theta_n,
\end{equation}
where $E^{(i)}_j$ is the electric field at output mode $j$ given the input at mode $i$. The interference visibility is then defined analogously to Eq.(3) in the main text,
\begin{equation}
V=\frac{P^{(\infty)}-P^{(0)}}{P^{(\infty)}}.
\end{equation}

\vspace{5mm}
\noindent{\bf IV. Effects of non-ideal photon sources from downconversion \vspace{5mm}}

\noindent Spontaneous parametric downconversion, as the name implies, is a probabilistic phenomenon which outputs photon states of the form
\begin{equation}
\label{eq:s1}
\left | \psi \right >  \propto \left | 00 \right >+ \eta \left | 11 \right >+ \eta^2 \left | 22 \right >+ \eta^3 \left | 33 \right > ...,
\end{equation}
where: $\left | n_a n_b \right >$ gives the photonic occupation numbers $n_a$ and $n_b$ in the spatial modes $a$ and $b$ respectively \cite{Broome:11}; the probability amplitude of creating one photon-pair per pump pulse is $\eta$, which incorporates the nonlinear interaction strength, interaction time of the pump field with the nonlinear crystal and the optical coupling efficiency of the down-converted modes, and $\eta{\ll}1$. Triggering using one photon of a pair removes the vacuum component, but since $\eta^{i}{>}0$ there is a possibility that two or more photons remain in the signal modes---these are the so-called higher-order terms. 

Higher-order terms can never be entirely removed from downconversion light with linear optics alone. Their detrimental effect can be reduced though by keeping $\eta$ at a necessary minimum, while keeping the single-photon rate constant through temporal \cite{Broome:11} or spatial \cite{ma2011egs} multiplexing of multiple downconversion sources. 

\end{document}